%% file: main.tex
\documentclass[runningheads]{llncs}
\usepackage{graphicx}
\usepackage{tabularx}

\usepackage{booktabs} 
\usepackage[normalem]{ulem}
\useunder{\uline}{\ul}{}

\usepackage{color}

\usepackage{graphicx}

\begin{document}
\title{Identifying Historical Travelogues in Large Text Corpora Using Machine Learning}
\titlerunning{Identifying Historical Travelogues}
%
\author{
	Jan R\"orden\inst{1}\orcidID{0000-0002-5824-8397} \and
	Doris Gruber\inst{2}\orcidID{0000-0002-0512-100X} \and
	Martin Krickl\inst{3}\orcidID{0000-0002-8569-0340} \and
	Bernhard Haslhofer\inst{1}\orcidID{0000-0002-0415-4491}
}

\authorrunning{J. R\"orden et al.}


\institute{
	AIT Austrian Institute of Technology, Vienna, Austria \and
	Austrian Academy of Sciences, Vienna, Austria \and
	Austrian National Library, Vienna, Austria
}

\maketitle              

\begin{abstract}
Travelogues represent an important and intensively studied source for scholars in the humanities, as they provide insights into people, cultures, and places of the past.
However, existing studies rarely utilize more than a dozen primary sources, since the human capacities of working with a large number of historical sources are naturally limited.
In this paper, we define the notion of \emph{travelogue} and report upon an interdisciplinary method that, using machine learning as well as domain knowledge, can effectively identify German travelogues in the digitized inventory of the Austrian National Library with F1 scores between 0.94 and 1.00.
We applied our method on a corpus of 161,522 German volumes and identified 345 travelogues that could not be identified using traditional search methods, resulting in the most extensive collection of early modern German travelogues ever created. To our knowledge, this is the first time such a method was implemented for the bibliographic indexing of a text corpus on this scale, improving and extending the traditional methods in the humanities. Overall, we consider our technique to be an important first step in a broader effort of developing a novel mixed-method approach for the large-scale serial analysis of travelogues.

\end{abstract}


\input{sections/introduction}

\input{sections/background}

\input{sections/methods}

\input{sections/results}

\input{sections/discussion}

\input{sections/conclusions}

\section*{Acknowledgments}


The work in the Travelogues project (\url{http://www.travelogues-project.info}) is funded through an international project grant by the Austrian Science Fund (FWF, Austria: I 3795) and the German Research Foundation (DFG, Germany: 398697847).

%
%
%
\bibliographystyle{splncs04}
\bibliography{bibliography}
\end{document}

%% file: sections/introduction.tex
\section{Introduction}\label{section:introduction}


Travelogues offer a wide range of information on topics closely connected to current challenges, including mass tourism, transnational migration, interculturality and globalization. By definition, documents considered to be travelogues contain perceptions of \emph{Otherness} related to foreign regions, cultures, or religions. At the same time, travelogues are strongly shaped by the socio-cultural background of the people involved in their production. Comparative analysis allows us, in turn, to scrutinize how (specific) cultures handled \emph{Otherness}, as well as to examine the evolution of stereotypes and prejudices. This high degree of topicality fosters the continuous growth of studies on travels and travelogues, as can be observed by the sheer flood of publications appearing every year (c.f.~\cite{salzani2010bibliography}). While heuristic approaches proved to be fruitful~\cite{AgaiConermann2013,vanGroesen2008}, many fundamental questions connected to travelogues remain unanswered, among other reasons, because previous analysis of travelogues rarely exceeded a dozen primary sources.


In response, we seek to leverage the possibilities offered by large-scale digitization efforts, as well as novel automated text-mining and machine learning techniques, for the first time, on travelogues. This allows us to significantly increase the quantity of text we can analyze. The overall goal of our work is to develop a novel mixed qualitative and quantitative method for the serial analysis of large-scale text corpora and apply that method to a comprehensive corpus of German language travelogues from the period 1500--1876 (ca. 3,000--3,500 books) drawn from the Austrian Books Online project (ca. 600,000 books) of the Austrian National Library (ONB).

As a first step, and this is the focus of this paper, we seek to provide automated support for scholars in identifying travelogues in large collections of historical documents, which have been scanned and undergone an optical character recognition (OCR) process by Google. A major challenge clearly lies in finding an effective method that can be scaled for large collections, is robust enough to support documents with varying OCR quality and can deal with the evolution of the German language over almost four centuries. Previous studies have already demonstrated the potential of quantitative methods for investigating cultural trends~\cite{michel2011quantitative} or types of discourses in the past~\cite{Purschwitz2018Netzwerke}, and the effectiveness of automated machine learning techniques for subject indexing~\cite{mai2018using}. However, to the best of our knowledge, no method has previously been tailored to the specific characteristics and unique challenges of identifying travelogues.

To this end, our contributions can be summarized as follows:

\begin{enumerate}

    \item We reviewed the characteristics and commonalities of travelogues and combine our findings into a generic definition of a \emph{travelogue}.

    \item We provided a manually annotated dataset of documents that match our working definition of a travelogue in the range of the 16th to the 19th century.\footnote{We will share the corpus here: \url{https://github.com/Travelogues/travelogues-corpus}.}

    \item We employed that dataset as a ground-truth for evaluating a variety of document classification methods and found that a multilayer perceptron (MLP) model trained with standard bag-of-words (BOW) and bag of n-grams (range 1, 2) feature set can effectively identify travelogues with an F1 ratio of 1 (16th c.), 0.94 (17th c.), 0.94 (18th c.) and 0.97 (19th c.).\footnote{The code (as Jupyter notebook) that we used for the classification is available here: \url{https://github.com/Travelogues/identifying-travelogues}.}

    \item We found that approximately 30 manually annotated documents are needed for training an effective classifier.

\end{enumerate}

Our results show that standard machine learning approaches can effectively identify travelogues in large text corpora. When we applied our most effective model on the ONB's entire German language corpus, we unearthed 345 travelogues that could not be identified using a traditional keyword search. Thus, we were able to create the most extensive collection of early modern German travelogues to date. This will provide us a solid baseline for determining subsequent steps to develop a serial text-analysis method, which will focus on the specific phenomena of intertextuality and analysis of semantic expressions referring to \emph{Otherness}.

We will present our definition of travelogue and closely related work in the next section. Afterward, in Section~\ref{section:methodology}, we outline our methodology before presenting our results in Section~\ref{sec:results}. Finally, we discuss the implications and limitations in Section~\ref{section:discussion} and conclude our paper in Section~\ref{sec:conclusions}.

%% file: sections/background.tex
\section{Background}\label{section:background}

\subsection{Characteristics of Travelogues}\label{subsection:travelogues-character}

For identifying travelogues we needed, first of all, a precise definition of the notion of a \emph{travelogue}. In previous research, very broad and general definitions were suggested that, unfortunately, did not resolve all of our questions connected to the classification~\cite{Piera2018Travel,Zimmermann2002}. There was, for instance, no conclusive answer whether or not missives, letters of consuls, or texts only partly including descriptions of actual travel are to be considered travelogues. Consequently, we had to generate our own definition, which aims to apply to all historical eras, geographical regions and media types. Our considerations, however, which are based on an analysis of printed (early) modern travelogues in German, have been formed accordingly and can be characterized as follows:

\begin{quote}
    A travelogue is a specific type of media~\cite{GenzGevaudan2016} that reports on a journey which, if detectable, actually took place. Consequently, a travelogue is formed by two relations: the first is content-based (description of a journey) and the second biographical (factuality of the journey).
\end{quote}

Our definition builds upon and refines the careful reflections of Almut H{\"o}fert~\cite{Hofert2003Feind}, who provides a narrower characterization: Fictional narratives are excluded, but there is no binary distinction between fictionality and factuality, since a certain amount of fictionality is part of every travelogue~\cite{Nunning2008Wirklichkeit,sandrock2015truth}, apparent factuality was often generated artificially, and fictional narratives influenced reports at times~\cite{Stagl2002Geschichte2}.

A journey is a movement in space and time that begins at a starting point and then moves through a variable set of further points outside of the well-known cultural environment of the traveler. In contrast to Wolfang Treue~\cite{Treue2014Abenteuer} we include (forced) emigrations and relics of people who died while traveling, but exclude movements on a permanent level (e.g., nomads, vagabonds).

Travelogues can be handed down in various forms, whether through oral speech, non-verbal communication, text, an image or video. Travelogues obtained from the (early) modern period and available for research consist of text and/or images. The available text is predominantly in prose and can be attributed to several text genres, such as reports, diaries, letters or missives. Notwithstanding that there are many mixed forms and transition zones here, especially since certain guidelines for the creation of travelogues (the so-called \emph{Ars Apodemica}) only emerged and were, if at all, partially applied by the authors during the course of the study period~\cite{kurbis2004hispania,Stagl1983Apodemiken}.

The only decisive element of a text to be classified as travelogue is the mention of the fact that it reflects the experiences of an actual journey that was undertaken, with all of the imaginable variations of spellings and semantic forms. A frequent, but not always included, feature is an itinerary listing different stations along the journey and the connected experiences associated with the stops. Images in travelogues are usually mimetic, predominantly including portraits, landscapes, and depictions of plants, animals or architecture, but may also incorporate abstract representations. The decisive element here is the inclusion of any pictorial form that is a reflection about experiences that occurred during a journey. Thus, a series of pictures, which originated from an actual journey and contain no text, are also understood to be travelogues, but are not collected within the current project that is focusing specifically on text.

Most of the travelogues from the (early) modern period were written by the traveling persons themselves, are therefore known as \emph{ego documents}~\cite{Presser1969Memoires}, and, predominantly, in a narrower sense considered to be \emph{self-testimonies} (Selbstzeugnisse)~\cite{Krusenstjern1994,Ludtke1993}. Consequently, the personal experiences and cultural background of the authors, as well as other persons involved in the production of the final document, strongly shape the content of the resulting texts.

However, (early) modern travelogues should not be considered detached from each other, since they depend on each other and/or other (types of) media intertextually~\cite{Pfister1993Intertext}, interpictorially~\cite{Greve2004Bild}, intermaterially or intermodally~\cite{BellingradtSalman2017}. For the definition itself it is considered irrelevant whether, in the case of a publication, a travelogue was published by the traveling person or by someone else (e.g., posthumous publications, later editions, written/edited by a related person), whether they are independent publications, appear in the context of a travel collection, as part of a larger publication (e.g., autobiography, historiography) or in the form of an excerpt.

\subsection{Known Document Identification Methods}

Generally, linear classifiers have demonstrated solid performance for text classification tasks. This includes support vector machines (SVM) and logistic regression, as shown in \cite{joachims1998text} and \cite{fan2008liblinear}. We build on these findings and evaluate both methods in our experiments.

A recent study in the digital library field, by Mai et al.~\cite{mai2018using}, compared the effectiveness of classification models trained on titles only versus models trained on full-texts and found that the former outperform the latter. They used multilayer perceptron (MLP), convolutional neural network (CNN) and long short-term memory (LSTM) architectures, and found that MLP outperformed the other methods in most cases. Although their models were trained on large-scale datasets from other domains (PubMed, EconBiz) and therefore not directly applicable, we consider MLPs for building a travelogue identification model.

Dai et al.~\cite{dai2017social} use an unsupervised method based on word embeddings to cluster social media tweets as related or unrelated to a topic, in their case influenza. Although they use much shorter texts (Twitter posts, or tweets, were limited to 140 characters until 2018), their task is similar to the one we present here in that both are binary classification tasks. The authors report an F1 score as high as 0.847, using pre-trained word embeddings from the Google News dataset. Additionally, the authors compared their approach to other methods such as keyword or related-word analysis but found their solution to perform better. In this paper, we will show that similar scores can be achieved without using pre-trained word embeddings.

In~\cite{yang2016hierarchical}, Yang et al. use hierarchical attention networks for document classification, in their case sentiment estimation and topic classification (multi-class). Their model outperforms previous methods, depending on the dataset, reaching F1 scores between 0.494 and 0.758. Additionally, they are able to visualize the informative components of a given document. This might be a suitable method for identifying possible subject indexing terms (classes) in an entire corpus and a subsequent document classification task. However, since the identification of travelogues is a binary classification problem, we refrain from these methods at the moment.

Zhang et al.~\cite{zhang2015character} use character-level convolutional networks for text classification, comparing them against methods such as BOW, n-grams and other neural network architectures. They test on several large-scale datasets (e.g., news, reviews, question/answers, DBPedia), showing that their methodology outperforms most of the other approaches, having up to 40\% fewer errors. While the authors do not report F1 scores, they illustrate that treating text as just a sequence of characters, without syntactic or semantic information or even knowing the words, can work well for classification tasks. While we do not apply their findings directly, we take inspiration from their work and use BOW and bag-of-n-grams features.

%% file: sections/methods.tex
\section{Methods}\label{section:methodology}

\subsection{Overview}\label{subsec:interdisc-method}

Our overall goal is to develop a novel mixed qualitative and quantitative method for the serial analysis of large-scale text corpora. Since serial analysis typically focuses on a specific topic or type of document, in this case \emph{travelogues}, we first need to define a systematic method that supports scholars with diverse backgrounds (historical science, library and information science, data science) in iteratively training a machine learning model that ultimately supports them in locating travelogues within a huge collection of digitized documents.

\begin{figure}
  \centering
  \includegraphics[width=\linewidth]{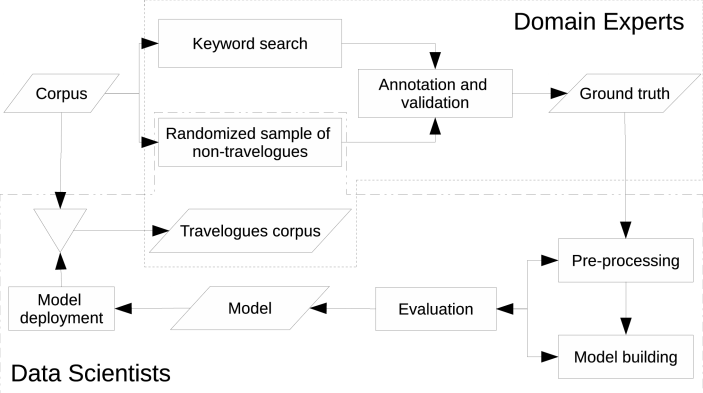}
  \caption{High-level overview of our interdisciplinary approach. \normalfont{Creation of the ground truth is primarily the responsibility of domain experts (with data scientists contributing to identify non-travelogues). Model creation is completed by data scientists, with the results of the model deployment on the whole corpus being evaluated by the domain experts again.}}
  \label{fig:approach-overview}
\end{figure}

Figure~\ref{fig:approach-overview} summarizes the overall workflow and involved participants from a high-level perspective: in the first step, domain experts use the keyword search feature of the Austrian National Library's catalog to search the \emph{overall corpus} for documents meeting our definition of a \emph{travelogue}. They manually inspect each result and annotate those matching our definition as being a \emph{travelogue}. In parallel, the data scientist automatically selects a randomized sample of documents from the overall corpus, which are then manually inspected and verified by the domain experts as being \emph{non-travelogues}. This process, which is described in more detail in Section~\ref{sec:dataset-collection}, yields a balanced \emph{ground truth} corpus consisting of travelogues and non-travelogues documents, which can then be used for subsequent machine learning tasks.

Before building machine learning models, documents in the ground truth corpus need to be pre-processed, which includes cleansing, normalization and feature engineering steps. Section~\ref{sec:preprocessing} explains in more detail the steps we applied to our documents. Next, we use the pre-processed documents for \emph{model building}, which includes training various machine learning models, such as SVMs and MLPs. This process is described in Section~\ref{sec:model_building}. Following this, we evaluate the effectiveness of the trained models (see Section~\ref{sec:evaluation}).

The top-performing model was then deployed and used to classify the remaining documents in our corpus, in an attempt to identify additional, potentially previously unknown travelogue documents. As a result, our iterative method yields a growing \emph{travelogue corpus}, which can be used for refining the effectiveness of machine learning tasks and for other quantitative and qualitative analytics tasks. In the following sections we describe each step in more detail.

\subsection{Dataset and Ground Truth Creation}\label{sec:dataset-collection}

In our work, we are focusing on prints published between 1500 and 1876, which are part of the historical holdings of the ONB. Since 2011, more than 600,000 books (volumes) from that period have been digitized and OCR-processed in a public-private partnership with Google (Austrian Books Online, ABO\footnote{\url{https://www.onb.ac.at/en/digital-library-catalogues/austrian-books-online-abo}.}). Therefore, nearly all of the library's historical books are currently accessible in a digital form. Within this corpus, we are specifically searching for travelogues.

As a first step, we identified German volumes in the overall ABO corpus, and then split the corpus by century. Then we initiated a ground truth by querying over titles and subject headings. We searched for different keywords in German, namely truncated spellings of `Reise' (travel) and `Fahrt' (journey) along with their known variants and with wildcard affixes and suffixes (in alphabetical order: *faart*, *fahrt*, *fart*, *rais*, *rai\ss{}*, *raisz*, *rays*, *ray\ss{}*, *raysz*, *reis*, *rei\ss{}*, *reisz*, *reys*, *rey\ss{}*, *reysz*, *rys*, *ry\ss{}*) as well as common subject-headings in the library's catalog including `Forschungsreise' (expedition), `Reise' (travel) and `Reisebericht' (travelogue).

As these queries still generated many false positives, we cleaned up the dataset manually. Results were double-checked intellectually by two annotators fluent in German and experienced with early modern German, a historian and a librarian, who read parts of the texts and utilized external bibliographies, biographies and catalogs,\footnote{E.g.: \url{https://www.deutsche-biographie.de/}, \url{https://lb-eutin.kreis-oh.de/},\\ \url{https://kvk.bibliothek.kit.edu/}, \url{https://www.oclc.org/de/worldcat.html},\\ \url{http://www.vd16.de/}, \url{http://www.vd17.de/}, \url{http://www.vd18.de/},\\ \url{https://viaf.org/}, Wikipedia.} to confirm whether a document meets our definition of travelogue or belongs to another genre. Uncertainties were resolved unanimously and there were no disagreements on the final annotations. The result of this step is a manually annotated and verified sample of travelogues, which took approximately three months of full time work for both annotators.

Since training and validation of machine learning models also requires counterexamples, in this case, non-travelogues, we implemented an automated procedure for randomly selecting an equally sized sample of documents from the subset of German volumes. Via a manual investigation process conducted by the same annotators, we ensured that those documents were not travelogues. This provides a manually verified sample of non-travelogues.

In total, our travelogues ground truth dataset contains a balanced sample of 6,048 volumes, representing 3.67\% of 167,570 German language volumes from the complete ONB corpus. Table~\ref{table:books-volumes} provides an overview of our ground truth and its distribution over centuries. One can easily observe that the number of volumes, as well as the size of each publication increases with time. To provide insight into how likely it is to find a travelogue randomly, we also included the number of travelogues that were found while reviewing the randomized sample we used as counterexamples. This approach was replicated for the 16\textsuperscript{th}, 17\textsuperscript{th}, 18\textsuperscript{th} and 19\textsuperscript{th} centuries. Volumes that were not evaluated remain in the \emph{candidates} pool, upon which we applied our classifier after identifying the best-performing model.

\begin{table}
\centering
\caption{Dataset overview. \normalfont Our corpus consists of the total number of digitized German-language books available to us. The ground truth contains an equal amount of travelogues and randomly selected counter examples; in brackets, we provide the number of travelogues we found by chance. Books not evaluated remain in the \emph{candidates} pool. A token contains at least two alphanumerical characters, punctuation etc. is not counted.}
\label{table:books-volumes}
\begin{tabular*}{\textwidth}{@{\extracolsep{\fill}}crrrr@{}}
\toprule
\multicolumn{5}{c}{\textbf{Corpus}} \\ \midrule
\multicolumn{1}{l}{Century} & \multicolumn{1}{c}{\begin{tabular}[c]{@{}c@{}}No. candidate\\ volumes\end{tabular}} & \multicolumn{1}{c}{\begin{tabular}[c]{@{}c@{}}No. ground\\ truth volumes\end{tabular}} & \multicolumn{1}{c}{Total tokens} & \multicolumn{1}{c}{Average tokens} \\ \midrule
16th & 8,526 & 67/67 & 362,244,353 & 41,829 \\
17th & 8,763 & 161/161 & 651,957,983 & 71,762 \\
18th & 55,971 & 873/873 & 5,041,741,840 & 82,274 \\
19th & 88,262 & 1,897/1,897 & 11,464,645,150 & 124,539 \\ \midrule
$ \sum $ & 161,522 & 5,996 & 17,520,589,326 & Ø 104,589 \\ \bottomrule
\end{tabular*}
\end{table}

\subsection{Pre-processing}\label{sec:preprocessing}

The preprocessing phase involves several steps. First, the texts were tokenized at the word level, using blanks and interpunctuation as separators. The German language uses upper- and lowercase spelling, depending on the word type and their position in the sentence, but to compensate for OCR and orthographic errors we transformed all tokens to lowercase. Furthermore, we removed all tokens that do not contain at least two alphanumeric characters, as this removes OCR artifacts, which are often special characters. For the same reason, each token needs to appear at least twice in the whole corpus.

\subsection{Model Building}\label{sec:model_building}

As shown in Table~\ref{table:books-volumes}, the documents that we seek to classify are rather large, as they contain on average 41,000-124,000 tokens. We decided to use a combination of BOW and bag-of-n-grams, as shown by (Wang and Manning,~\cite{wang2012baselines}). With this approach, we can both handle intricate problems with our data, while having a computationally effective method that still provides competitive results.

Experiments were performed on the above-mentioned ground truth. We tested different classification algorithms:

\begin{itemize}
    \item Multinominal Naive Bayes (MNB)
    \item Support Vector Machine (SVM)
    \item Logistic regression (Log)
    \item Multilayer perceptron (MLP)
\end{itemize}

For the MNB, SVM and Log algorithms we used the sklearn~\cite{scikit-learn} implementation. We use the Tensorflow~\cite{tensorflow2015-whitepaper} and Keras~\cite{chollet2015keras} implementation for the MLP. The data for all algorithms was vectorized and hashed with the sklearn HashingVectorizer.

\subsection{Evaluation Procedure}\label{sec:evaluation}

As a baseline, we applied a random classification. In all the experiments, we treat every book as a single document.

First, we split the ground truth into both a training (75\%) set and a validation (25\%) set, for every time period. We evaluated all classifiers presented here first through a five-fold cross evaluation along the training split. This essentially means that the training set was split into five equally sized subsets, and for each fold one subset serves as a test, and the other four become the training data. When the results across the cross-evaluation are comparable, good scores are less likely to occur by chance. Subsequently, we applied the classifiers on the held-out validation data. The classification results are discussed in Section~\ref{sec:results}.

The evaluation of our work follows a two-step approach. First, we gauge the effectiveness of a given method by precision, recall and F1 metrics on our training set. \emph{Precision} is the number of correct results, divided by the number of all returned results. \emph{Recall} shows how many of the documents that should have been found are actually found, dividing the number of correctly classified documents by the number of documents that actually belong to that class. \emph{F1} is the harmonic mean of precision and recall (with a range between 0 and 1, with 1 representing the perfect result).

For the second step of our evaluation, we apply the model that performs best on our training data to the remaining documents of our corpus. This results in a list of all those documents, with probability scores indicating how likely they belong to our travelogues class. Starting with the highest probability, those documents are then manually evaluated by domain experts (see Section~\ref{sec:results}), to judge how well this model identifies travelogues in a set of unseen documents.

\subsection{Minimal Ground Truth Requirements}\label{sec:clf-efficiency}

Additionally, we wanted to understand how many ground truth documents are needed to train an effective classifier. We approached this by testing the top classification approach against different amounts of ground truth documents and evaluated the results. The same setup as described above is used, but we varied the number of ground truth documents, as well as randomizing their selection.

For each time frame, we evaluated 5, 10, 15, 20, 25, 30 and 50 examples each for the positive class (travelogue) and the negative class (anything else); for the 18\textsuperscript{th} and 19\textsuperscript{th} centuries, we extended to 100 examples each. The model created with those documents was then tested on the remaining ground truth documents. For each sample size, we repeated this a total of five times with a different randomized sample each time.

%% file: sections/results.tex
\section{Results}\label{sec:results}

\subsection{Classification Results}

Table~\ref{tabel:results} shows the evaluation of our classification algorithms.

\begin{table}
\centering
\caption{Classification results. We provide precision, recall and F1 scores for multinominal naive Bayes (MNB), support vector machine (SVM), logistic regression (Log) and a multi-layer perceptron neural network (MLP).}
\label{tabel:results}
\begin{tabular*}{\textwidth}{@{\extracolsep{\fill}}c|lll|lll|lll|lll@{}}
\toprule
\multicolumn{1}{l|}{} & \multicolumn{3}{c|}{MNB} & \multicolumn{3}{c|}{SVM} & \multicolumn{3}{c|}{Log} & \multicolumn{3}{c}{MLP} \\ \midrule
Century & \multicolumn{1}{c}{P} & \multicolumn{1}{c}{R} & \multicolumn{1}{c|}{F1} & \multicolumn{1}{c}{P} & \multicolumn{1}{c}{R} & \multicolumn{1}{c|}{F1} & \multicolumn{1}{c}{P} & \multicolumn{1}{c}{R} & \multicolumn{1}{c|}{F1} & \multicolumn{1}{c}{P} & \multicolumn{1}{c}{R} & \multicolumn{1}{c}{F1} \\
16th & 0.73 & \textbf{1.00} & 0.85 & 0.96 & \textbf{1.00} & 0.98 & 0.95 & 0.95 & 0.95 & \textbf{1.00} & \textbf{1.00} & \textbf{1.00} \\
17th & 0.75 & 0.97 & 0.84 & 0.82 & 0.92 & 0.87 & 0.82 & 0.92 & 0.87 & \textbf{0.95} & \textbf{0.93} & \textbf{0.94} \\
18th & 0.79 & \textbf{0.94} & 0.86 & 0.88 & 0.90 & 0.89 & 0.84 & 0.88 & 0.86 & \textbf{0.96} & 0.93 & \textbf{0.94} \\
19th & 0.86 & 0.92 & 0.89 & 0.91 & 0.90 & 0.91 & 0.88 & 0.91 & 0.90 & \textbf{0.97} & \textbf{0.96} & \textbf{0.96} \\ \bottomrule
\end{tabular*}
\end{table}

Our results show that it is possible to achieve good classification scores with our dataset, even without extensive feature engineering or pre-trained word embeddings. We assume that this is based in the comparably\footnote{Many works focus on datasets that have more, but shorter documents, c.f.~\cite{yang2016hierarchical} for comparisons of multiple classification methods and datasets.} large size of our data points, as can be seen in Table~\ref{table:books-volumes}. Additionally, through the randomness involved in the selection of non-travelogues, those volumes are expected to have a high variance in genres, which matches the whole corpus as well. Comparing numerous examples of one genre against an equal number of volumes covering many more genres certainly benefits the classification, especially when taking into account the length of the documents.


Taking the results from this evaluation, we were confident in approaching our main task, which was the identification of travelogues from a much larger dataset: the other digitized books in our corpus not yet evaluated by us.

Following the training of models suitable for classification, one specifically designed for each century, we applied it to our pool of \emph{candidates}. We used this process to create a list of books that are potentially travelogues, ranked from highest to lowest, and evaluated the first 200 items. The results of this are shown in Table~\ref{tabel:discovered_travelogues}, and have been subject to the same scrutiny as our initial ground truth.

We can show that our methodology proves a clear improvement over a less guided evaluation: within our evaluated findings, true positives made up 12.5\% (16\textsuperscript{th} c.), 30\% (17\textsuperscript{th} c.), 41.5\% (18\textsuperscript{th} c.) and 89.5\% (19\textsuperscript{th} c.) respectively. Due to time constraints, our evaluation was discontinued after the first 200 items, but this already means that we discovered 345 books of the travelogue genre that were not found by traditional search queries on meta data, as we explained in Section~\ref{sec:dataset-collection}. Discovery by chance only resulted in 3\% (16\textsuperscript{th} c.) and 0.8\% (18\textsuperscript{th} c.) positives, or none at all (17\textsuperscript{th} c., 19\textsuperscript{th} c.).

It also has to be noted that the increase in the percentage of true positives in a diachronic perspective is connected to the equally increasing number of travelogues. There simply are many more travelogues that were printed in the 19\textsuperscript{th} century than in the previous time periods.

\begin{table}
\centering
\caption{Applying the classifier on the candidates pool. \normalfont \emph{By chance} shows how many travelogues were found when randomly selecting examples for the ground truth.}
\label{tabel:discovered_travelogues}
\begin{tabular*}{0.8\textwidth}{@{\extracolsep{\fill}}crrr@{}}
\toprule
\textbf{Century} & \textbf{\begin{tabular}[c]{@{}c@{}}No. \\ candidates\end{tabular}} & \multicolumn{1}{c}{\textbf{\begin{tabular}[c]{@{}c@{}}Confirmed\\ (top 200)\end{tabular}}} & \textbf{By chance} \\ \midrule
16\textsuperscript{th} & 8,526 & 25 & 2 (from 67) \\
17\textsuperscript{th} & 8,763 & 60 & 0 (from 161) \\
18\textsuperscript{th} & 55,971 & 83 & 7 (from 873) \\
19\textsuperscript{th} & 88,262 & 177 & 0 (from 1,897) \\ \bottomrule
\end{tabular*}
\end{table}

\subsection{Error Analysis}

During this evaluation, it became apparent that the majority of potential findings with a high probability belong to the group we attribute as historiography. Due to their nature, they have strong a overlap with travelogues but lack the required criteria of describing a journey actually experienced by the author. This crucial information can in many cases only be gathered from external sources\footnote{C.f. Section \ref{subsection:travelogues-character}.}. From a purely technical perspective, there is no difference with the content of the books between travelogues and historiographies. This means that while, for the purposes of this project, they are very different, right now we cannot further differentiate between them. Additionally, in the 18\textsuperscript{th} century, many false positives include publications on geography; a possible explanation here is that they often describe locations, which naturally overlap with travelogues as well.

\subsection{Ground Truth Requirements}\label{subsec:efficiency-results}

\begin{figure}
  \centering
  \includegraphics[width=\linewidth]{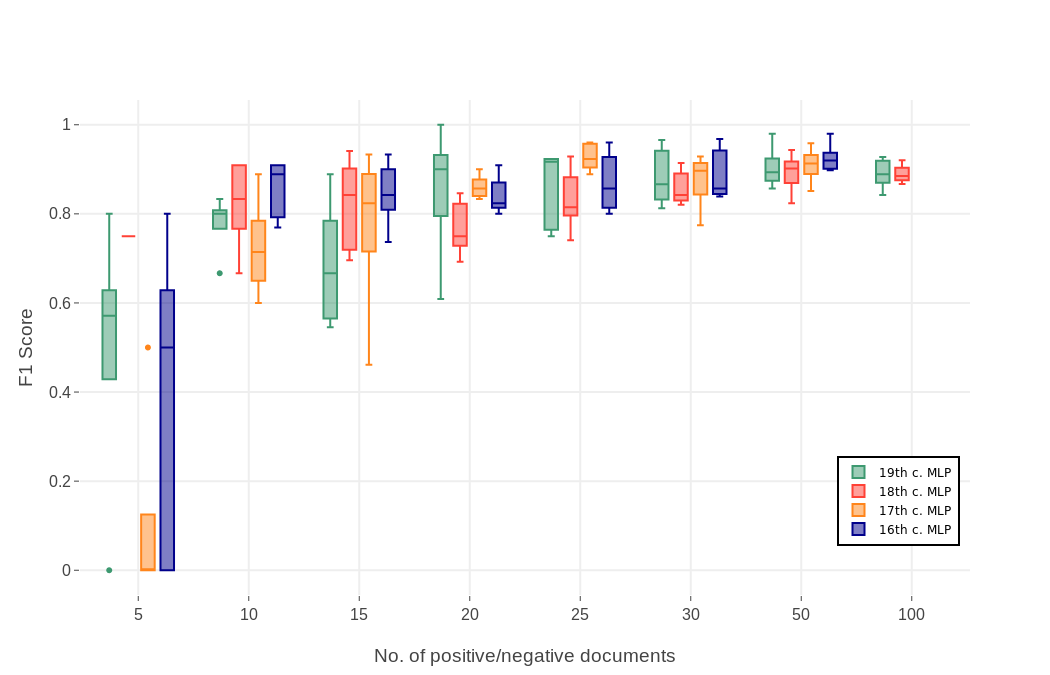}
  \caption{Classifier efficiency evaluation for MLP. Every step has a balanced number of travelogues and non-travelogues (5/5, 10/10, 15/15, 20/20, 25/25, 30/30, 50/50 and 100/100). The experiment was repeated five times for each step.}
  \label{fig:efficiency-plot}
\end{figure}

The result of our efficiency evaluation is depicted in Figure~\ref{fig:efficiency-plot}. We provide the average F1 score for each time frame, as well as the variance between the samples.

For every time frame, our general observation is that the performance of the MLP classification fluctuates heavily when only a small dataset is available. With at least 20 documents, but it is better with 30 examples, each for the positive and negative class a stable performance of above a 0.8 F1 score can be reached. After that it slowly increases up to a 0.9 F1 score at 50 examples each, with only very minor changes for 100 examples each.

This experiment shows that it is possible to create a working classification methodology, which reaches acceptable results with a modest time investment\footnote{Depending on the sources, and if additional definitions etc. are needed, between several hours and up to a few weeks of full time work.} upfront, as shown in Table~\ref{tabel:results}.

%% file: sections/discussion.tex
\section{Discussion}\label{section:discussion}

Our results show that standard machine learning techniques combined with relatively easily computable features (BOW and bag-of-n-grams) can effectively support scholars in identifying travelogues in a large-scale document corpus. Using the same features for an MLP neural network generates even superior results. This is an important first step in the development of a broader mixed-method approach for the large-scale serial analysis of travelogues.

Specifically, we discovered a total of 345 travelogues in the evaluated time periods using the top 200 findings with the highest confidence scores each (800 in total). We previously were not able to find any of these files through search queries based on meta data or manual search in our catalog, hence this directly translates into a re-discovery of sources for scholars in the humanities.

Additionally, a large fraction of false positives is, at the level of words and their semantics, extremely similar to the true positives. However, going by the definition provided earlier we are required to use external information that could not be included as a feature, as it is dependant on domain knowledge. This severely limits the efficacy of unsupervised machine learning and deep learning approaches.

A clear limitation of our effort lies in the time and effort required to create a high-quality ground truth. While this effort could possibly be reduced by applying unsupervised clustering techniques beforehand, annotations provided by domain experts will always be key for effective learning techniques. Applying active learning techniques (c.f.~\cite{settles2012active}) for iteratively developing ground truths could be a possible strategy for reducing this manual annotation effort. Another limitation of our approach lies in the focus on entire volumes, which currently neglects the fact that volumes may include travelogues and non-travelogues. Using a wider spectrum of semantically richer features (c.f.~\cite{momeni2013identification}) such as named entities could support classification at the paragraph or page level.

Nonetheless, our experiments on the efficiency of the classification method presented here show that it is possible to achieve robust results above an F1 score of 0.8 with a relatively small ground truth size. Knowing this, future research in different domains should require substantially less time investments to get started.

A remaining challenge lies in the distinction between highly similar genres, in our case historiographies or geographic books and travelogues. We hope that this can be tackled by further refining the ground truth to fit the given genres, taking into account a wide range of external sources, to include domain knowledge in a structured way.

%% file: sections/conclusions.tex
\section{Conclusions}\label{sec:conclusions}

In this paper, we have described a methodology to identify historical travelogues in a large dataset. Our approach combines the knowledge of both domain experts (historical science, library and information science) and data scientists to create a ground truth and subsequently build an MLP model, successfully identifying 345 previously unknown travelogues. Furthermore, we have shown that a ground truth for this kind of data can be as small as 30 examples each for the positive and negative class and still perform well.

In the upcoming weeks and months, we will begin looking at the discovered travelogues in more detail. In a first step, we will identify intertextual relations in our corpus to find out in what way the travelogues depended on each other, why and how (certain) stereotypes and prejudices were handed down, evolved or disappeared over the centuries.

Ultimately, we want to know how foreign cultures, places and people were perceived, and if the perceptions differed depending on the socio-cultural background of the involved people. This will allow us to come to an understanding how and why something was perceived as \emph{Other}, if and how this changed across the centuries. For this, we have done the groundwork here, as it is crucial to rely on as much data as possible; concrete next steps in this direction include creating a formal description of intertextuality and \emph{Otherness}, and translating it into a set of machine-readable text features.